# Power Law Spectra in the Nature: Analogies with the Cosmic Ray Spectrum

Yu. V. Stenkin
*Institute for Nuclear Research of Russian Academy of Sciences*
*60th October anniv. prospect 7a, 117312 Moscow, RUSSIA*
Presenter: Yu.V.Stenkin (stenkin@sci.lebedev.ru), rus-stenkin-Y-abs3-he12-poster

Examples of the power law spectra observations in the Nature are given. It is shown that the observed meteoroid mass spectrum is very similar to the observed cosmic ray energy spectrum. It is concluded that the "knees" observed in both spectra have a similar origin: the Earth's atmosphere. Both spectra are studied by indirect methods through secondary components and specific behavior of these components in the atmosphere could produce the "knee" even in a case when primary spectrum follows a pure power law.

## 1. Introduction

Power law spectra are widely-distributed spectra in the Nature. They can be found among absolutely different objects in different sciences from mathematics [1] to biology, economics etc. [2]. As it has been shown by P.Bak [3], the author of the theory of self-organized criticality (SOC), such behavior is natural for Nature: "The theory of self-organized criticality posits that complex behavior in nature emerges from the dynamics of extended, dissipative systems that evolve through a sequence of meta-stable states into a critical state, with long range spatial and temporal correlations. Minor disturbances lead to intermittent events of all sizes." It is interesting that power-law is a scale-free function and thus, it could be expected in all processes with a scaling or fractal behavior. *"Self-organized critical systems"* (SOC) evolve toward a scale-free, or critical state naturally, without fine tuning any parameters. This gives rise to power law distributions for the breakdown events. Minimal SOC models have been developed to describe a diverse set of phenomena including earthquakes, solar flares, forest fires, magnetically confined plasma, fluctuations in stock-markets and economics, black hole accretion disks, traffic, biological evolution, braided rivers formed by vortex avalanches in superconductors, and disease epidemics, among others." [2].

Application of this theory to cosmic rays can also give rather interesting consequences.

## 2. Meteoroid mass distribution

The most interesting for us example of a quasi-power-law spectrum is the measured spectrum of meteoroid masses. It is interesting by several reasons: i) similar to Extensive Air Shower (EAS) observational methods (for meteors); ii) similar media of observations (Earth's atmosphere) and iii) similar results. The modern method of meteors observation is very similar to our luminescent method of EAS study. Both methods use an array of light-detecting devices, both methods record the light produced in the Earth's atmosphere and both methods are indirect: they record secondary component (light). The latter means that it is necessary to make complicated calculations to recover the primary spectrum. Fig. 1 shows meteoroid mass distribution measured or calculated in very wide range of masses by various methods and by various research groups [4]. As one can see the measured spectrum reminds the cosmic ray spectrum. It also has a "knee" at mass $m \sim 3$ Kg and an "ankle" at mass $m \sim 10^7$ Kg. Note that expected power-law spectrum with integral index=-0.833 [5] at the top of atmosphere is shown by the upper dot-dash straight line. Experimental points shown on the graph are not recalculated to primary spectrum at the top of atmosphere as in cosmic ray physics people usually do. Therefore, experimental points represent here spectra of different secondary components such as recovered photometric mass of the biggest fragment (in a case of meteors), estimated mass of a meteorite able to produce such a crater etc. The spectrum around the "knee" was obtained by the meteor observation method. Let us consider this method in details.

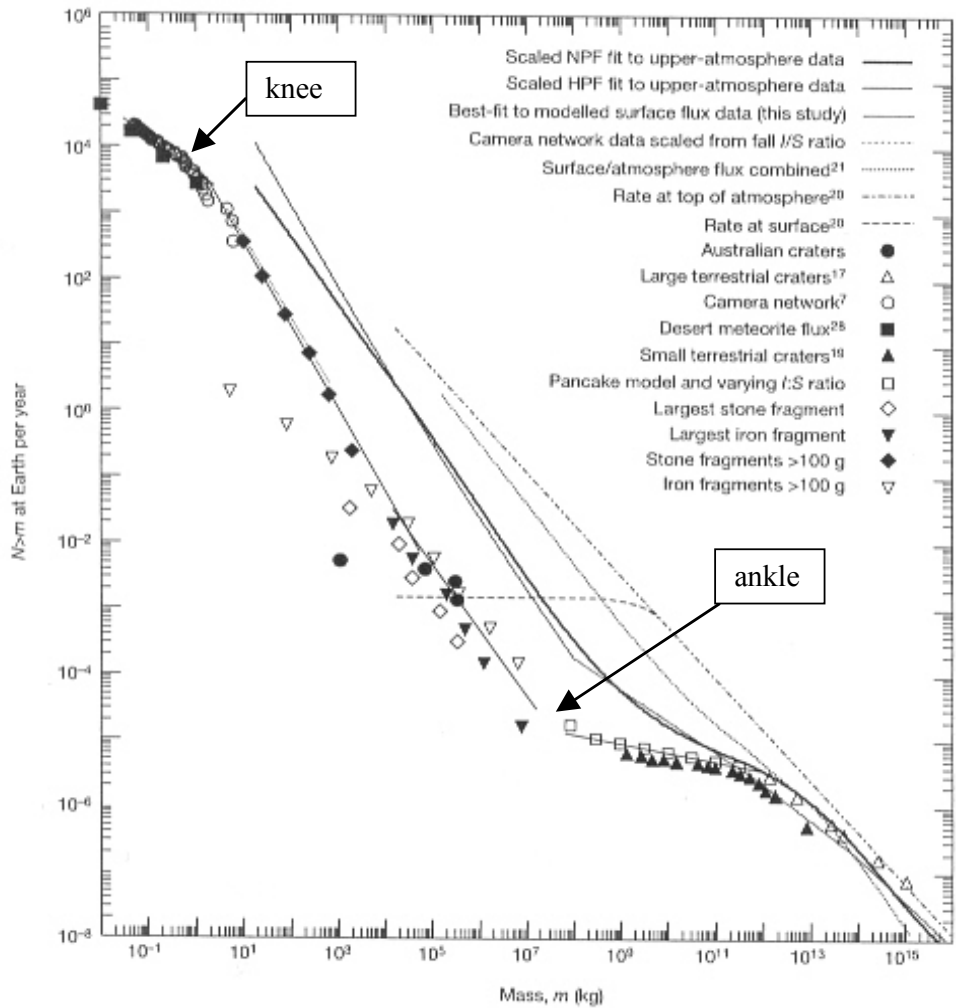

**Figure 1.** Meteoroid mass spectrum [4].

The most complete results on this subject are presented by the Canadian camera-network of Meteorite Observation and Recovery Project (MORP) [6, 7]. It covers ~$10^6$ Km$^2$ with cameras distributed over all Canada. Their results are shown in Fig.2 as fits of experimental data for various meteoroid types. Upper curve (labeled all) shows expected total flux. This curve has a clear visible knee at MI~3 Kg. But, the authors never said that it characteristic of primary mass distribution. Applying the " survival probability" [7] to the expected power-law spectrum [5] they obtained a good agreement with measured mass MI distribution (MI is a sum of photometric mass of the biggest fragment + estimated total mass of survived meteorite fragments). Using "survival probability" of meteoroids in the atmosphere from [7] we plotted in Fig.3 the dependence of measured mass on initial mass at the top of atmosphere.

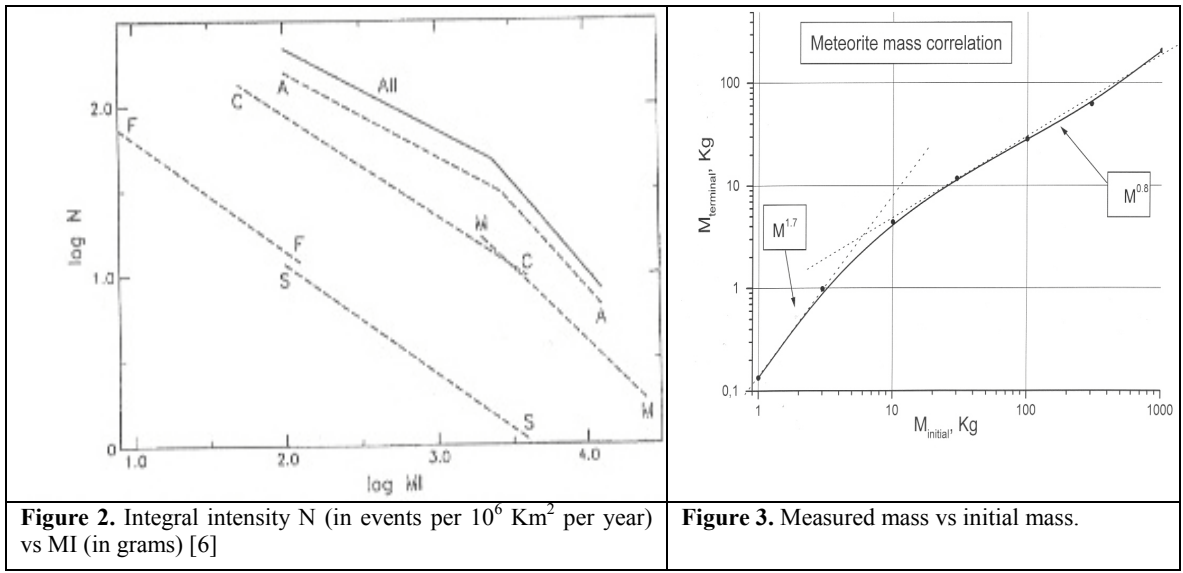

**Figure 2.** Integral intensity N (in events per $10^6$ Km$^2$ per year) vs MI (in grams) [6]

**Figure 3.** Measured mass vs initial mass.

This plot shows how the secondary mass (terminal or measured MI) depends on primary mass (initial, before entering Earth's atmosphere). This curve has also a kink around 3 Kg. A reason for such a behavior is obvious: meteoroids of masses higher than ~3 Kg always reach the Earth's surface while those with smaller masses do not. For them absolutely another parameters help to survive, such as initial velocity, chemical composition etc. That is why the survival curve has a kink. This results in a "knee" in observed spectrum $I(m_x)$ because two power-law indices affect it: $I(m_x) \sim m_x^{-\gamma/\alpha}$ when primary spectrum is $I(m) \sim m^{-\gamma}$ and secondary component depends on primary one (m) as: $m_x \sim m^{\alpha}$.

## 3. Analogy with the cosmic ray spectrum

A similar situation could be found in cosmic ray spectrum study. Due to very low intensity of cosmic rays of energy above 1 PeV only indirect methods based on the EAS method are used to study their energy spectrum. As it was mentioned above, EAS luminescent and/or Cherenkov light detection methods have the closest analogy with meteors study. Analogy could also be found in physics of both phenomena. In both cases primary objects must have big enough kinetic energy to penetrate thick atmosphere and reach the Earth's surface. Initial velocities of meteoroids do not differ significantly and their penetrating ability depends mostly on their mass. As we saw above this threshold mass is equal to ~3 Kg. For relativistic primary cosmic ray particles it depends on their energy per nucleon. Similar to meteoroid case, primary particle (nucleus) fragmentizes quickly to constituent nucleons. According to the *superposition principle* each nucleon interacts independently and gives a start of its own hadronic cascade. Altogether they produce EAS. Hadrons play a crucial role in EAS development. Similar to meteoroid fragments they can reach the Earth's surface only if the energy of initial nucleon is high enough. It has been shown [8] that this threshold energy is equal to ~ 100 TeV/nucleon at sea level. Above this energy there exists equilibrium (due to decays of neutral pions and kaons)) between cascading hadrons and electromagnetic EAS components: electrons, gammas and Cherenkov light. Below this threshold the equilibrium is broken and EAS looks (and developing) like a pure electromagnetic shower with addition of muons. Again as in the case of meteoroids, absolutely another parameters affect the EAS development here and this results in a kink (see Fig.1 in [8]) in the dependence of EAS size ($N_e$) as a function of primary energy ($E_0$) similar to that shown in Fig.3. It is easy to see that primary proton produces this "knee" at $E_0 \sim 100$ TeV while primary iron only at $E_0 \sim 5$ PeV.

Due to very small amount of primaries heavier than iron in cosmic rays, there should be seen 2 "knees" in all-particle EAS size spectrum at these 2 points even in a case of pure power-law primary spectrum. (This effect has no analogy with meteoroids because superposition principle is not valid for them.) As a result observed "knee" at $E_0 \sim 5$ PeV appears in this model naturally in the needed place and of the needed size [8]. If in recovering primary spectrum one uses a fit for calculated dependence of $N_e(E_0) \sim E_0^\alpha$ with fixed $\alpha$ than visible "knee" will be assigned to primary spectrum.

## 4. Conclusion

Power-law spectra are widely distributed in Nature due to self-organized criticality principle. This allows one to hope that primary cosmic ray spectrum also follows power law. A theory of cosmic ray acceleration in cosmic plasma pinches [9] gives such a spectrum with index $\gamma = \sqrt{3} \approx 1.73$ very close to that measured by direct methods. Analogy with meteoroid mass spectrum clearly shows how indirect measurements could produce a "knee" in a case when primary spectrum follows pure power law.

***

Author thanks Román Pérez-Enríquez for discussion and provided references.
This work was partially supported by RFBR grant 05-02-17395.